\documentclass[%
 reprint,
superscriptaddress,
 amsmath,amssymb,
 aps,
floatfix,
]{revtex4-2}

\usepackage{graphicx}
\usepackage{dcolumn}
\usepackage{bm}
\usepackage{color}

\UseRawInputEncoding

\let\oldphi\rho 
\let\rho\varrho 
\let\varrho\oldphi

\begin{document}

\preprint{APS/123-QED}

\title{Phonons behave like Electrons in the Thermal Hall Effect of the Cuprates
}

\author{Liuke Lyu}
\affiliation{Département de Physique, Université de Montréal, Montréal, Québec, H3C 3J7, Canada}

\author{William Witczak-Krempa}
\affiliation{Département de Physique, Université de Montréal, Montréal, Québec, H3C 3J7, Canada}
\affiliation{
 Institut Courtois, Université de Montréal, Montréal (Québec), H2V 0B3, Canada
}
\affiliation{
 Centre de Recherches Mathématiques, Université de Montréal, Montréal, QC, Canada, HC3 3J7
}

\date{\today}

\begin{abstract}
The thermal Hall effect, which arises when heat flows transversely to an applied thermal gradient, has become an important observable in the study of quantum materials. Recent experiments found a large thermal Hall conductivity $\kappa_{xy}$ in many high-temperature cuprate superconductors, including deep inside the Mott insulator, but the underlying mechanism remains unknown. Here, we uncover a surprising linear temperature dependence for the
inverse thermal Hall resistivity, $1/\rho_H=-\kappa_{xx}^2/\kappa_{xy}$, in the Mott insulating cuprates $\mathrm{La_2CuO_4}$ and $\mathrm{Sr_2CuO_2Cl_2}$. We also find this linear scaling in the pseudogap state of Nd-LSCO in the out-of-plane direction, highlighting the importance of phonons. On the electron-doped side, the linear inverse thermal Hall signal emerges in NCCO and PCCO at various dopings, including in the strange metal. Although such dependence arises in the simple Drude model for itinerant electrons, its origin is unclear in strongly correlated Mott insulating or pseudogap states. We perform a Boltzmann analysis for phonons that incorporates skew-scattering, and we are able to identify regimes where a linear $T$ inverse Hall resistivity appears. Finally, we suggest future experiments that would further our fundamental understanding of heat transport in the cuprates and other quantum materials.

\end{abstract}

\maketitle


\textit{Introduction}---The thermal Hall effect occurs when a system is subjected to a temperature gradient that gives rise to a flow of heat in the transverse direction, in analogy to the electrical Hall effect. Due to its ability to detect neutral excitations, it has recently been used to probe quantum spin liquid candidate materials~\cite{Ong,Yamashita}, and to provide experimental evidence for the elusive Majorana fermions in topological states of matter that are insulating in their bulk~\cite{Stern,Matsuda}.
Subsequent experiments found an unexpected large thermal Hall effect in a wider variety  of insulators, such as the cuprate Mott insulators $\mathrm{La_2CuO_4}$ and $\mathrm{Sr_2CuO_2Cl_2}$, the quantum paraelectric $\mathrm{SrTiO_3}$, and the cubic antiferromagnetic insulator $\mathrm{Cu_3TeO_6}$~\cite{Grissonnanche2019,Boulanger2020,Li2020,LuChen2021}. The temperature dependences of the longitudinal thermal conductivity $\kappa_{xx}$ and transverse (or Hall) thermal conductivity $\kappa_{xy}$ are often shown; with their ratio $\kappa_{xy}/\kappa_{xx}$ quantifying the relative magnitude of the thermal Hall effect. 
Recent theoretical works~\cite{mangeolle2022,Guo2022} have suggested that the \emph{thermal Hall resistivity} $\rho_H \equiv -\kappa_{xy}/\kappa_{xx}^2$ is often simpler to interpret than $\kappa_{xy}/\kappa_{xx}$. 
Physically, the thermal Hall resistivity $\rho_H = \nabla_y T / J_x$ represents the transverse temperature gradient due to a given longitudinal heat flux.
One motivation to study this quantity comes from the observation that extrinsic mechanisms for the thermal Hall effect usually predict:
\begin{equation}
    \kappa_{xx} \sim \tau, \quad \kappa_{xy} \sim  \tau^{2}
\end{equation}
where $\tau$ is the total relaxation time, which thus cancels out in $\rho_H$ (see, for example,~\cite{XiaoQi2022}). 
Since $\tau$ can be sample-dependent and  difficult to model, it is advantageous to study a quantity independent of $\tau$. In this work, we find that the inverse thermal Hall resistivity $\rho_H^{-1}$ possesses a simple temperature dependence in a wide variety of cuprates. 

We first begin by showing that the experiments for the cuprates show a linear inverse thermal Hall resistivity $\rho_H^{-1}\approx A\, T$ in various undoped, hole-doped, and electron-doped materials. We also find that  $\rho_H$  scales linearly with the applied field for an electron-doped cuprate.  We then explain how such temperature and field scaling appears in the simple Drude model for itinerant electrons. Given the inapplicability of such a model to the cuprates, we study a kinetic Boltzmann equation for heat carriers, such as phonons. Under certain assumptions that we motivate, we uncover a linear scaling in agreement with experiments.
We end with a discussion of the implications of our findings, outlining future experiments in various materials to investigate the conditions for such striking linear scaling.
\begin{figure}[ht]
\includegraphics[scale=0.53]{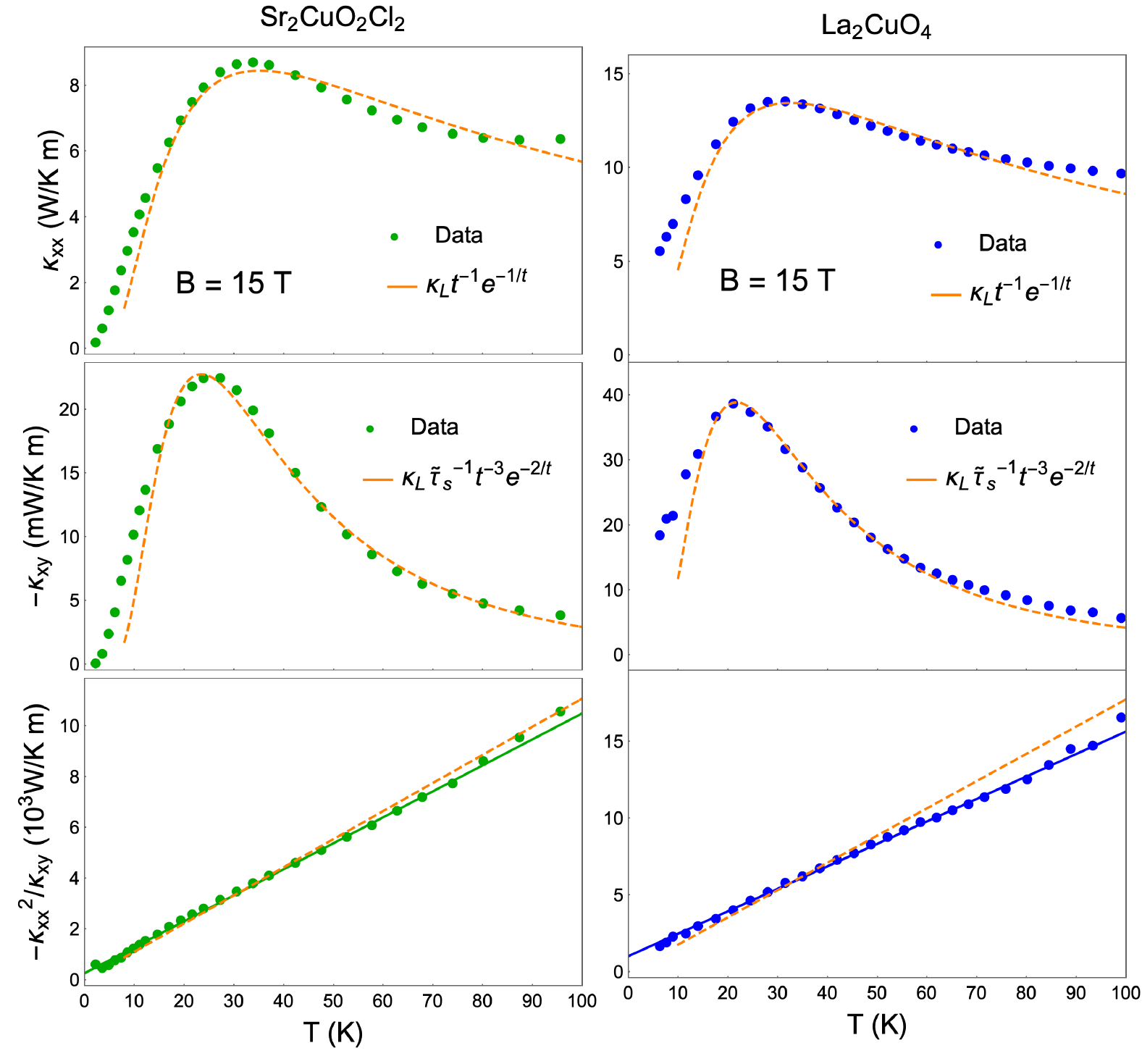}
\caption{\label{fig:mott} Thermal longitudinal conductivity (top), Hall conductivity (middle), and inverse thermal Hall resistivity $\rho_H^{-1}=-\kappa_{xx}^2/\kappa_{xy}$ (bottom) for the cuprate Mott insulators SCOC and LCO. 
The dashed lines show the approximate behavior of the skew scattering model in the intermediate temperature regime, where $t=T/(a\Theta_D)$ is $T$ rescaled by the Debye temperature and a dimensionless factor $a$ (see (\ref{eq:kinetic})).
The thick lines in the bottom show a linear fit, $AT+A_0$, with $A^{\mathrm{Sr/La}}= 0.10, 0.15 \mathrm{\times 10^3 W/(K^2 m)}$ , $A_0^{\mathrm{Sr/La}}= 0.25, 0.95 \mathrm{\times 10^3 W/(K m)}$.
Data from~\cite{Boulanger2020}.
}
\end{figure}

\emph{Thermal Hall resistivity of the cuprates}---
We now examine the quantity $\rho_H$ in various cuprates, using data from recent experimental works~\cite{Boulanger2020, Grissonnanche2020chiral, Boulanger2022}.
The first group of materials includes the undoped cuprate Mott insulators $\mathrm{La_2CuO_4}$ (LCO) and $\mathrm{Sr_2CuO_2Cl_2}$ (SCOC) with a similar crystal structure and antiferromagnetic order. 
Both materials contain layered copper-oxygen planes, where the $\mathrm{Cu^{2+}}$ moments form a 2D antiferromagnet with a Néel temperature $T_N\approx 300$~K~(La) and $250$~K~(Sr), respectively~\cite{Vaknin1990}. 
Below $530$~K, $\mathrm{La_2CuO_4}$ has an orthorhombic structure with spins slightly canting out of the $\mathrm{CuO_2}$ plane, while $\mathrm{Sr_2CuO_2Cl_2}$ remains in a tetragonal structure down to at least $10$~K~\cite{Vaknin1990}.
We found that $\rho_H^{-1}$ scales linearly with $T$ over a wide range of temperatures.
\begin{equation} 
    \rho_H^{-1} = -\frac{\kappa_{xx}^2}{\kappa_{xy}} \approx A\, T + A_0 \label{eq:ratio}
\end{equation}
as shown in Fig.~\ref{fig:mott}. 
This is remarkable given that the longitudinal and transverse thermal conductivities possess non-monotonic temperature dependence (Fig.~\ref{fig:mott}), with maxima at different temperatures. This shows that a hitherto undiscovered correlation exists between $\kappa_{xx}$ and $\kappa_{xy}$. 

\begin{figure}[ht]
\centering
\includegraphics[scale=0.32]{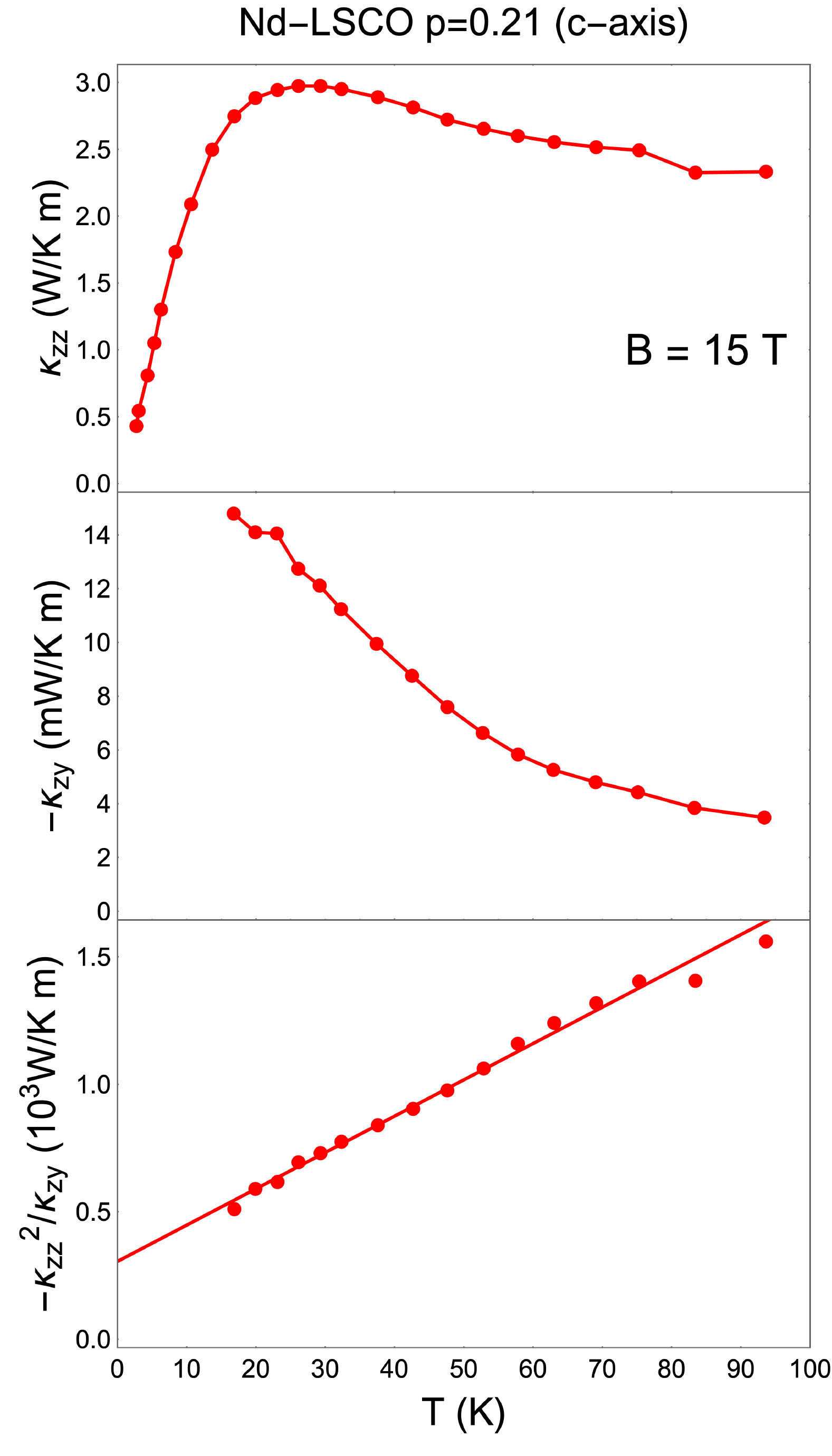}
\caption{\label{fig:Nd-LSCO} Thermal conductivities (top and middle), and $\rho_H^{-1}$ (bottom) along the $c$-axis versus temperature for the hole-doped cuprate Nd-LSCO at doping $p=0.21$. 
The linear fit corresponds to $A=0.014 \mathrm{\times 10^3 W/(K^2 m)}$, and $A_0=0.32 \mathrm{\times 10^3 W/(K m)}$.
Data from~\cite{Grissonnanche2020chiral}.
}
\end{figure}

In order to better understand the role of phonons in the linear scaling (\ref{eq:ratio}) compared to electronic excitations, including magnons, we now examine thermal transport along the $c$-axis. We find that the linear scaling (\ref{eq:ratio}) is also present in the pseudogap phase of $\mathrm{La}_{1.6-x}\mathrm{Nd}_{0.4}\mathrm{Sr}_{x}\mathrm{CuO}_4$ (Nd-LSCO) at hole doping $p=0.21$.
As shown in Fig.~\ref{fig:Nd-LSCO}, the $c$-axis inverse thermal Hall resistivity is linear in a wide range of temperatures above 20~K. 
This shows that the linear behavior (\ref{eq:ratio}) is present not only beyond the Mott insulating regime but also perpendicular to the CuO planes, pointing to the central role of phonons in the effect. 
We note that the magnitude of $\rho_H^{-1}$ is smaller by a factor of 10 compared to the undoped cuprates. 

\begin{figure}[ht]
\centering
\includegraphics[scale=0.55]{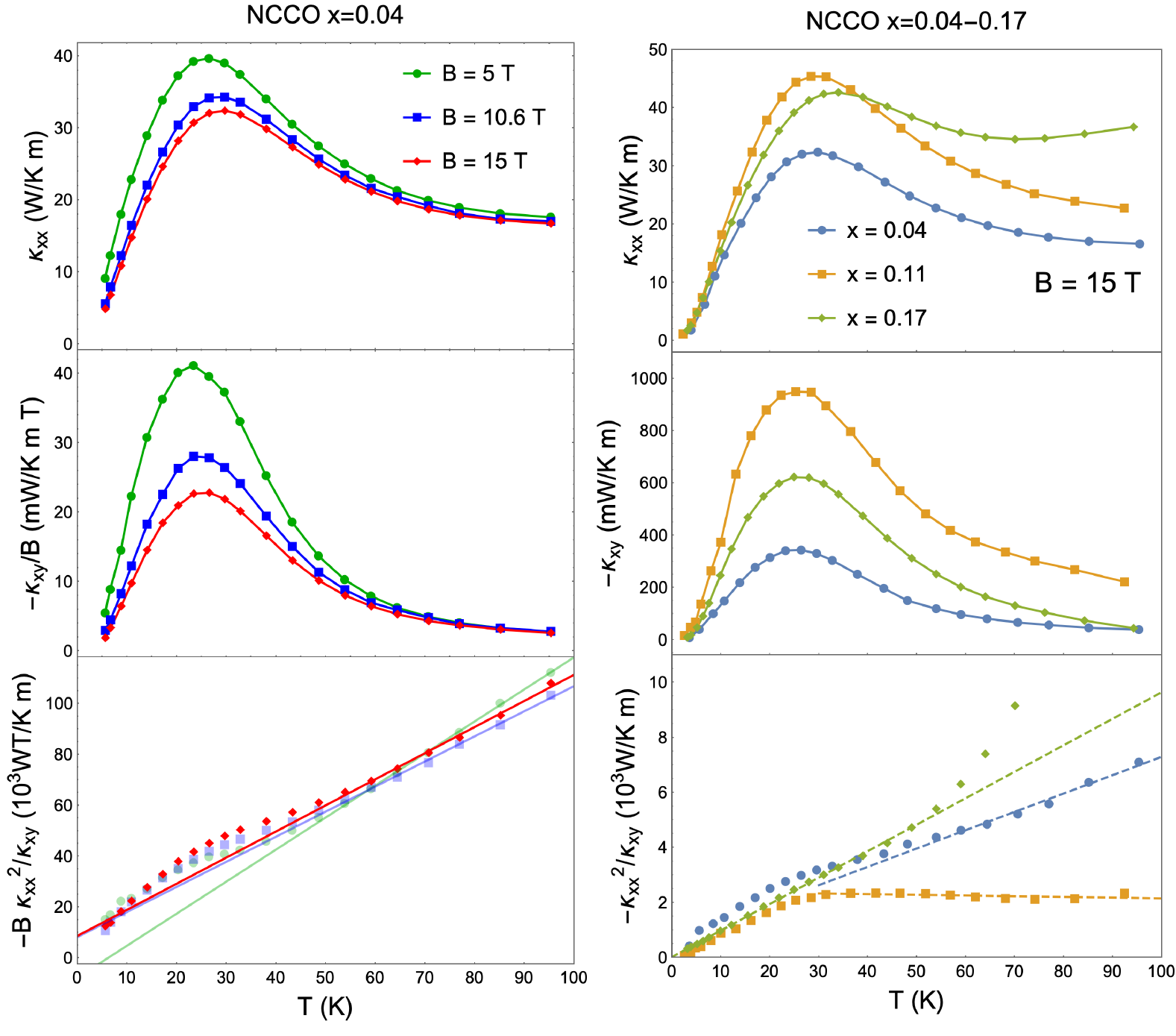} 
\caption{\label{fig:NCCO} Thermal conductivities (top and middle), and $\rho_H^{-1}$ (bottom) versus temperature for the electron-doped cuprate NCCO at doping $x=0.04$ (left) and 0.17 (right).
In the left column, where $T\geq 5$~K, we see that $\kappa_{xy}/B$ strongly depends on the applied field $B$, but the inverse thermal Hall coefficient $R_{\text{TH}}^{-1} = B/\rho_H$ shows a much weaker field dependence. 
The linear fit for $B$ = 5, 10.6, 15~T gives $A$ = 0.26, 0.09, 0.07 $\mathrm{\times 10^3 W/(K^2 m)}$, and $A_0$ = -1.56, 0.77, 0.57 $\mathrm{\times 10^3 W/(K m)}$.
On the right column, the linear fits for $x$ = 0.04, 0.11, 0.17 corresponds to $A$=0.07, -0.003, 0.10 $\mathrm{\times 10^3 W/(K^2 m)}$, and $A_0$ = 0.6, 2.4, 0.02$ \mathrm{\times 10^3 W/(K m)}$.
Data from~\cite{Boulanger2022}. 
} 
\end{figure}

Now we turn to electron-doped cuprate $\mathrm{Nd}_{2-x}\mathrm{Ce}_x\mathrm{CuO}_4$ (NCCO).
At doping $x=0.04$, the sample is in the antiferromagnetic Mott insulating phase. 
We see that both $\kappa_{xx}$ and $\kappa_{xy}/B$ are suppressed with increasing field  (Fig.~\ref{fig:NCCO} left, top/middle). 
The curves $\kappa_{xy}/B$ vary significantly with the field, while the inverse thermal Hall coefficient $R_{\text{TH}}^{-1}=B \rho_H^{-1}$ (Fig.~\ref{fig:NCCO} left bottom) remains nearly the same for the three field values, and increases linearly with temperature beyond $50$~K.
In that range, $\rho_H^{-1}$ thus shows a $T/B$ dependence; we will see below that dependence also appears in the Drude model. 
Next, we analyze the doping dependence of thermal Hall resistivity in NCCO at $x$=0.04, 0.11 and 0.17. At doping $x$=0.11, the sample is still in the AF phase.
At $x$=0.17, which is above optimal doping, the sample is in the superconducting phase at zero magnetic field with $T_c = 6$~K~\cite{Helm2009}.   
Upon applying a $15$~T magnetic field, superconductivity disappears and leaves behind a non-Fermi liquid~\cite{Lambacher2008}.
We see a linear scaling beyond 50~K for the two samples in AF phase, and in particular at $x$=0.11 the slope almost vanishes.
For the $x$=0.17 metallic phase,  $\rho_H^{-1}$ scales linearly with $T$ until $60$~K, where it starts deviating from linearity. 
Part of the deviation could be an experimental artifact resulting from the black-body radiation of the sample~\cite{bb-radiation-private}. 
Going from $x$=0.11 to $x$=0.17, we see a decrease in $\kappa_{xy}$ and an increase in the slope of $\rho_H^{-1}$, possibly due to the vanishing magnetic order beyond critical doping, showing the importance of magnetic correlations in the thermal Hall effect. 
On the other hand, from doping $x$=0.04 to 0.11 within the AF phase, we see an opposite change in these quantities.
The doping dependence can be understood if we have an impurity-dependent microscopic mechanism, an example of which would be skew scattering off magnetic defects.
Impurities distort the AF magnetic order locally, generating an effective field that scatters phonons chirally.
Doping increases the concentration of impurities, creating more magnetic texture in the sample which leads to more frequent chiral scattering events. As magnetic correlations weaken upon approaching the quantum critical point, this effect becomes less effective. 
The combination of magnetic order and local distortions of this order create a chiral environment or effective Berry curvature~\cite{Zhang_2016} for phonons, which can generate a sizable thermal Hall effect. 

$\mathrm{Pr}_{2-x}\mathrm{Ce}_x\mathrm{CuO}_4$ (PCCO) is another electron-doped cuprate closely related to NCCO. 
At doping $x=0.15$ and zero field, PCCO becomes superconducting at $T_c\approx 20$~K~\cite{Takagi1989}.
In the metallic phase at $15$~T,  
the inverse thermal Hall resistivity (see Fig.~\ref{fig:PCCO} in Appendix B) shows a temperature dependence similar to NCCO, but the deviation from linearity occurs at a lower temperature $\sim 40$~K. 
Using the electrical Hall conductivity, an estimate of the electronic contribution $\kappa_{xy}^e$ was obtained by assuming the Wiedemann-Franz law~\cite{Boulanger2022}. 
At low temperature ($T < 40$~K), $\kappa_{xy}^e$ is seen to be of opposite sign and much smaller in magnitude than the measured thermal Hall conductivity (less than 10\%).
Before moving to the theoretical description, we note that the slope $A$ in the electron-doped cuprates is of similar magnitude to what was found for the undoped Mott insulators LCO and SCOC, $A\sim 10^{2}\,\mathrm{ W/(K^2 m)}$.

\textit{Drude model}---Before turning to a more realistic description of the thermal transport in the cuprates, we point out that the simple Drude model for itinerant electrons (or holes) shows this exact temperature dependence for the thermal Hall resistivity (\ref{eq:ratio}).
For an electron gas in a uniform magnetic field $B$, the Drude longitudinal and transverse electrical conductivities are 
\begin{equation}\label{eq:drude}
    \sigma_{xx}=\frac{nq^2\tau}{m}, \quad \sigma_{xy} = -\omega_c\tau \sigma_{xx},
    \quad \frac{\sigma_{xx}^2}{-\sigma_{xy}} = \frac{nq}{B}
\end{equation}
where  $\omega_c=qB/m$ is the cyclotron frequency, $q$ is the electric charge, $n$ is the charge carrier number density, $m$ is the effective mass, and $\tau$ is the relaxation time.
A key observation is that $-\sigma_{xx}^2/\sigma_{xy}$  is independent of $\tau$ and thus independent of temperature. Now, the Wiedemann-Franz law relates the thermal and electrical conductivities: $\kappa=L\sigma T$, where $L=\frac{\pi^{2}}{3}(k_B/e)^{2}$ is the Lorenz number, so that we can relate the thermal ratio (\ref{eq:ratio}) to the corresponding one for charge conductivities:
\begin{equation} \label{eq:WF}
    \rho_H^{-1} = \frac{\kappa_{xx}^{2}}{-\kappa_{xy}} = \frac{(L\sigma_{xx} T)^{2}}{-L\sigma_{xy}T}=L\frac{nq}{B}\, T 
\end{equation}
We thus recover a linear in $T$ inverse thermal Hall resistivity. We note in passing that to obtain the same sign as in Figs.~1-3, the carriers should be holes ($q>0$).
However, the arguments leading to (\ref{eq:WF}) cannot be applied to Mott insulators where the electric charge is localized at low temperatures. 
The dominant heat carriers in cuprate Mott insulators ought to be phonons, and to a lesser extent magnons. 
Phonons acquire chirality through either intrinsic mechanisms which are independent of impurities, or extrinsic ones which result from skew scatterings off impurities and defects.
Previous work has argued that the intrinsic phonon Hall effect due to phonon Berry curvature is at least $10^{-4}$ times smaller than the experimental results~\cite{Chen2020}, while others showed a complementary intrinsic effect due to phonons scattering off magnons are roughly comparable to experimental data~\cite{mangeolle2022}. 
Some recent work discussed the possibility of chiral transport by phonons due to skew scattering~\cite{XiaoQi2022, Guo2021}, and we now examine how it can give rise to the linear relation (\ref{eq:ratio}).

\textit{Skew Scattering Model}---We begin with the Boltzmann equation for bosonic heat carriers:
\begin{multline}\label{eq:Boltzmann}
  \frac{\partial f_{\mathbf{k}}}{\partial\omega_{\mathbf{k}}}\omega_{\mathbf{k}} \frac{\mathbf{v}_{\mathbf{k}}\cdot \nabla T}{T} =
  -\frac{\delta f_{\mathbf{k}}+\delta f^{\prime}_{\mathbf{k}}}{\tau(\omega_{\mathbf{k}})} \\
  +\int d^3 \mathbf k^{\prime}\left(W_{\mathbf{k}^{\prime} \mathbf{k}}^{A} \delta f_{\mathbf{k}}-W_{\mathbf{k} \mathbf{k}^{\prime}}^{A} \delta f_{\mathbf{k}^{\prime}}\right)
\end{multline}
where $\omega_{\mathbf{k}},\mathbf{v}_{\mathbf{k}}$ are the heat carrier frequency and velocity, respectively. $f_\mathbf{k}$ is the non-equilibrium heat carrier distribution, which we divide into three components: $f_{\mathbf{k}}=\bar{f}_\mathbf{k}+\delta f_\mathbf{k}+\delta f^{\prime}_\mathbf{k}$. The first term $\bar{f}$ represents the equilibrium distribution, which does not carry current. The second term is the longitudinal perturbation $\delta f \propto \nabla_x T$ driven by the applied temperature gradient, taken to be along $x$. The third term is a skew distribution $\delta f^{\prime} \propto B \nabla_x T$ proportional to both the temperature gradient and the magnetic field.
We will assume that $W_{\mathbf{k}\mathbf{k}^{\prime}}^{A}$ results from elastic scatterings under a static potential which breaks time-reversal symmetry, which can be expressed in the form~\cite{Chen2020}: $W_{\mathbf{k}\mathbf{k}^{\prime}}^{A}= \Omega \hat{z}\cdot(\hat{k}\times\hat{k}^{\prime})\delta(\omega_{\mathbf{k}}-\omega_{\mathbf{k}^{\prime}})$, where $\Omega\hat z$ is a ``Berry curvature''-term proportional to the applied magnetic field, taken to be along $z$.
Since $W_{\mathbf{k}\mathbf{k}^{\prime}}^{A}\neq W_{\mathbf{k}^{\prime}\mathbf{k}}^{A}$ breaks detailed-balance, 
it is claimed that the collision kernel in (\ref{eq:Boltzmann}) should not involve a Bose factor $(1+f_{\mathbf{k}} )$ at least for static potentials~\cite{Sturman_1984}, while other authors argued that the Bose factor is needed to describe dynamical defects~\cite{XiaoQi2022}. We have made the former choice and do not include a Bose factor.
We now introduce the skew scattering rate into mode $\mathbf{k}$:
\begin{align}
    \tau_{s,\mathbf{k}}^{-1}&=\int d^{3}\mathbf{k}^{\prime}\left(\hat{v}_{\mathbf{k}}\cdot\hat{y}\right)W_{\mathbf{k}\mathbf{k}^{\prime}}^{A}\left(\hat{v}_{\mathbf{k}^{\prime}}\cdot\hat{x}\right) \nonumber \\
    &= \frac{k^{2}\Omega(k)}{v_\mathbf{k}} \int_{S^{2}}d^{2}\hat{k}^{\prime}\left(\hat{v}_{\mathbf{k}}\cdot\hat{y}\right)\hat{z}\cdot (\hat{k}\times\hat{k}^{\prime})\left(\hat{v}_{\mathbf{k}^{\prime}}\cdot\hat{x}\right)
\end{align}
where the integral is over the unit sphere. The thermal Hall conductivity is then
\begin{equation}\label{eq:kxy}
    \kappa_{xy}=-\int \frac{d^{3}\mathbf{k}}{(2\pi)^3}\, \hbar v^{2}\frac{\omega^{2}}{T}\frac{\partial \bar{f}}{\partial\omega}\tau^{2}\tau_{s,\mathbf{k}}^{-1}
\end{equation}
We will assume a minimal form for the relaxation time $\tau$ which includes boundary, impurity and Umklapp scatterings:
$\tau^{-1}(\omega, T)=\frac{v}{L}+A \omega^4+B T \omega^2 \exp \left(-\frac{\Theta_{\mathrm{D}}}{b T}\right)$, where $\Theta_{D}$ is the Debye temperature. 
Previous works on the skew scattering model have made either order-of-magnitude estimates for the conductivities or focused on the low-temperature limit where boundary scattering is dominant~\cite{XiaoQi2022, Guo2022}, but this is not enough to explain the linearity of the experimental data. 
We have instead focused on the intermediate temperatures 30-80~K and examined the behavior of $\rho_H^{-1}$ with different forms of $\tau_{s,\mathbf{k}}^{-1}$. 
Surprisingly, a linear regime emerges when $\tau_s^{-1}$ is independent of both temperature and frequency, similar to the cyclotron frequency $\omega_c$ in the Drude model. 
This constant skew scattering rate leads to a deviation of $\rho_H^{-1}$ from linearity at low $T$, but captures the essential behavior of $\kappa_{xx}$, $\kappa_{xy}$, and $\rho_H^{-1}$ in a wider intermediate temperature range for LCO and SCOC (see Fig.~1, and Fig.~\ref{fig:fullmodel} in Appendix~\ref{appendix:Skew Scattering model}). 
From this we can obtain the constant skew scattering rates $\tau_s^{-1}=1.7\times10^{-6} \mathrm{s^{-1}}$  and $2.7\times10^{-6} \mathrm{s^{-1}}$ for LCO and SCOC, respectively.
To extract the linear scaling from the integral expressions of $\kappa_{xy}$ and $\kappa_{xx}$, we make approximations between 30-80~K.
In this intermediate temperature range, we can infer from the slow decrease of $\kappa_{xx}$ that Umklapp scattering is not dominant, since we do not see the exponential freeze-out factor $\exp \left(\frac{\Theta_D}{b T}\right)$. 
Our parameter fitting also shows the relative magnitudes of different scattering rates: $\tau_{\text{bound}}^{-1} \ll \tau_{\text{imp}}^{-1}(\omega) \sim \tau_{\text{Umk}}^{-1}(\omega,T)$ for a typical phonon with frequency $\omega\sim k_B T/\hbar$. 
Taking these assumptions, $\kappa_{xx}$ and $\kappa_{xy}$ can be approximated by the simple forms: 
\begin{equation}\label{eq:kinetic}
   \kappa_{x x}=\kappa_L \frac{e^{-1 / t}}{t}, \;\; \kappa_{x y}=-\kappa_L  \tilde{\tau}_s^{-1} \frac{e^{-2 / t}}{t^3}, \;\;
   \frac{\kappa_{x x}^2}{\kappa_{x y}}=\kappa_L \tilde{\tau}_s t
\end{equation}
where $t=T/(a\Theta_D)$, $\tilde{\tau}_{s}^{-1}=\frac{e}{4}\frac{\tau_{s}^{-1}}{v/L}$ is the dimensionless skew scattering rate, $\kappa_L= \frac{k_{B}\omega_{D}^{3}}{6\pi^{2}v}\frac{L}{v}a^{3}$ is the magnitude of $\kappa_{xx}$, $a=(A\omega_{D}^{4}\frac{L}{v})^{-1/4}$ is a dimensionless factor inversely related to the impurity parameter $A$, $\omega_D$ is the Debye frequency.
With only three independent fitting parameters $\kappa_L,\tilde{\tau}_s,a$ to fit both $\kappa_{xx}$ and $\kappa_{xy}$, the approximate formula agrees reasonably well with the experimental data (see Fig.~\ref{fig:mott}).
The thermal Hall resistivity given by the approximate equation (\ref{eq:kinetic}) scales linearly with the magnetic field from our assumption $\tau_{s}^{-1}\propto \Omega \propto B$, so we obtain the same scaling $\rho_H\propto B/T$ as in the Drude model.
This is in agreement with the $B$-linear scaling observed for NCCO, Fig.~\ref{fig:NCCO}. 

\textit{Discussion}---
To summarize, we have discovered a linear temperature dependence of the inverse thermal Hall resistivity in a set of undoped, electron-doped, and hole-doped cuprates which exhibit a large thermal Hall effect. In the case of Nd-LSCO, linearity holds when the heat gradient is applied orthogonally to the CuO planes, in line with previous claims that chiral phonons are responsible for the thermal Hall effect in cuprates~\cite{Grissonnanche2020chiral}.
Although the well-known Drude model of itinerant electrons obeys the linear relation (\ref{eq:ratio}), it cannot be applied to the cases we studied, which range from the Mott insulating phase to the pseudogap.
Instead, we found that a Boltzmann transport equation for phonons (and other bosonic excitations), under reasonable conditions, gives rise to the linearity of $\rho_H^{-1}$.
It also yields a $1/B$ dependence with respect to the magnetic field, so we have $\rho_H^{-1}\propto T/B$.
An important step forward would be to identify the microscopic origin of the skew-scattering collision term in the Boltzmann equation and to better understand the role of other excitations such as magnons. 

Interestingly, the linear dependence over a wide range of temperatures (\ref{eq:ratio}) is not present in all Mott insulators. 
For the cuprate $\mathrm{Nd_2CuO_4}$, the linearity is present up to $20$~K, at which point $\rho_H^{-1}$ reaches a maximum and increases until reaching a second linear regime at $70$~K (see Fig.\ref{fig:Nd2CuO4} in Appendix B). The deviation from the linearity observed in LCO and SCOC, Fig.\ref{fig:mott}, could arise from additional scattering in the longitudinal direction, which can be seen from a sharper decrease of $\kappa_{xx}$. 
Compared to the other two Mott insulators, $\mathrm{Nd_2CuO_4}$ has certain distinct features.
First, unlike LCO and SCOC, the Nd ions in $\mathrm{Nd_2CuO_4}$ lack apical anions.
Second, $\kappa_{xx}$ is more sensitive to the external field.
This may result from additional spin-phonon scattering, which is more prominent in $\mathrm{Nd_2CuO_4}$ due to the large magnetic moment carried by the $\mathrm{Nd^{3+}}$ sites~\cite{Li-ballistic-magnon}.
$\mathrm{Nd_2CuO_4}$ also has spin reorientations between successive CuO layers at 30~K and 70~K both without and with field~\cite{SKANTHAKUMAR1989,Li2005}, which roughly matches the range where $\rho_H^{-1}$ deviates from linearity. 
To examine this effect, future experiments may compare $\mathrm{Nd_2CuO_4}$ with its sister compounds such as $\mathrm{Pr_2CuO_4}$ and $\mathrm{Sm_2CuO_4}$ which do not show such spin reorientations.


A large thermal Hall conductivity was also found in the mixed state of $\mathrm{YBa_2Cu_3O}_{6+x}$ (YBCO) at $x=1$ below $T_c$ = 89~K, which was attributed to asymmetric scattering of quasiparticles by pinned vortices~\cite{Krishana1995}. 
The magnitude of the Hall effect was roughly $\kappa_{xy}/\kappa_{xx}\sim 1\%$, which is comparable with the NCCO data in Fig.~\ref{fig:NCCO}. 
It was claimed that $\kappa_{xy}$ was dominated by the quasiparticles. 
Given the new understanding of chiral phonons in the cuprates, it would be desirable to revisit the thermal Hall effect in YBCO
as a function of doping, both in- and out-of-plane, in particular, to see whether the thermal Hall resistivity possesses a linear-$T$ regime. Interestingly, a $T$-linear relation was observed for the inverse \emph{electrical} Hall coefficient $R_H^{-1}$ above $T_c$ at various dopings in YBCO~\cite{Jin1998}. It remains unclear whether this bears any relation to the thermal case, Eq.~(\ref{eq:ratio}).  

On the experimental side, our work motivates the analysis of thermal transport in more quantum materials to understand the universality of the linear inverse thermal Hall resistivity. In particular, iridate materials such as $\mathrm{Sr_2IrO_4}$ have similar properties to LCO, so it would be interesting to see if they also possess a linear scaling (\ref{eq:ratio}) regime.
In addition, this linear scaling gives a simple yet stringent constraint on theoretical models.


\textit{Acknowledgment.}---
We thank L.\/ Taillefer and G.\/ Grissonnanche for key suggestions and comments. We also thank R. Boyack for helpful discussions, and 
L.\ Chen and M.-E.\ Boulanger for providing the experimental data. 
 This research was funded by a Team Research Project from FRQNT, a Discovery Grant from NSERC, a Canada Research Chair, and a grant from the Fondation Courtois.

\bibliography{thermal-resistivity.bib}

\widetext
\clearpage
\appendix
\section{Skew Scattering Model and parameter fittings for $\mathrm{La_2CuO_4}$ and $\mathrm{Sr_2CuO_2Cl_2}$}\label{appendix:Skew Scattering model}
\subsection{Fitting parameters for the skew scattering model}
\begin{figure}[ht]
\centering
\includegraphics[scale=0.75]{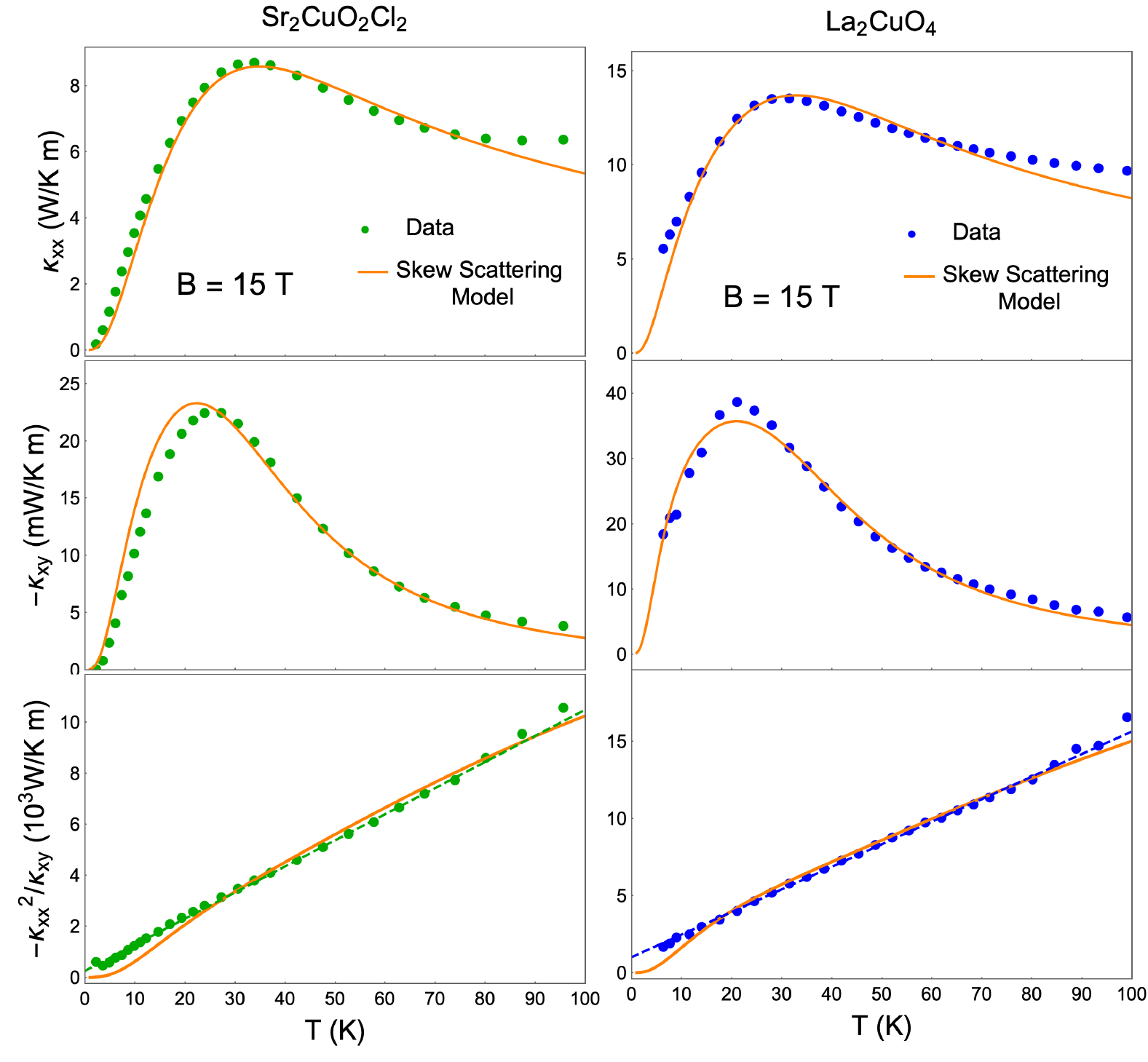} 
\caption{\label{fig:fullmodel} 
The full skew scattering model after parameter fittings on LCO and SCOC. We see a clear linear behavior of $\rho_H^{-1}$ between 30-80~K, and a deviation from linearity below 20~K. 
The approximate formula in Fig.~\ref{fig:mott} behaves essentially the same as the full model in the intermediate temperature regime. 
Data from~\cite{Boulanger2020}.
} 
\end{figure}

In the skew scattering model, $\kappa_{xx}$ and $\kappa_{xy}$ can be written as integrals in frequency space
\begin{align}
    \kappa_{xx}&=\frac{1}{3}v^{2}\int_{0}^{\omega_{D}}d\omega C(\omega)\tau(\omega,T)\\\kappa_{xy}&=\frac{1}{3}v^{2}\int_{0}^{\omega_{D}}d\omega C(\omega)\tau^{2}(\omega,T)\tau_{s}^{-1}(\omega)
\end{align}
where we define $C(\omega)=\frac{1}{2\pi^{2}v^{3}}\frac{1}{T}\omega^{4}\left(-\frac{\partial f_{B}}{\partial\omega}\right)$ with boson distribution function $f_B$. 
The longitudinal scattering rate is given by a combination of boundary scattering, impurity scattering, and Umklapp scattering processes:
\begin{equation}
    \tau^{-1}(\omega,T)=\frac{v}{L}+A\omega^{4}+BT\omega^{2}e^{-\frac{\Theta_{D}}{bT}}
\end{equation}
where $L$ is the Casimir length which is proportional to the size of the sample, $A$ is the impurity scattering coefficient and $B,b$ are the Umklapp scattering coefficients. 
We can put the integral in dimensionless form by defining dimensionless temperature and scattering rates: $t= T/\Theta_{D}$ (different from the definition in (\ref{eq:kinetic}) by a factor of $a$),
$\tilde{\tau}(x)=\frac{v}{L}\tau(\frac{k_{B}T}{\hbar}x)$, $\bar{\tau}_{s}(x)=\frac{v}{L}\tau_{s}(\frac{k_{B}T}{\hbar}x)$.
We can also define the thermal conductivity constant with boundary scattering alone $\kappa_{0}\equiv\frac{k_{B}}{6\pi^{2}v}\left(\frac{k_{B}\Theta_{D}}{\hbar}\right)^{3}\frac{L}{v}$. 
Furthermore, in our model, we have assumed $\tau_s^{-1}$ to be a constant of temperature and frequency, so we can pull it out of the integral. The final dimensionless form is
\begin{align}
\kappa_{xx} &=\kappa_{0}t^{3}\int_{0}^{1/t}dx J_4(x)\tilde{\tau}(x,t)\\
\kappa_{xy} &=\kappa_{0}\bar{\tau}_{s}^{-1}t^{3}\int_{0}^{1/t}dxJ_4(x)\tilde{\tau}^{2}(x,t)
\end{align}
where $x=\hbar\omega/k_B T$, are dimensionless frequency and temperature, and
\begin{equation}
    J_n(x)=\frac{x^{n}e^{x}}{(e^{x}-1)^{2}}
\end{equation}
The longitudinal scattering rates can also be put in a dimensionless form:
\begin{equation}
    \tilde{\tau}^{-1}(x,t)=1+\tilde{A}x^{4}t^{4}+\tilde{B}t^{3}e^{-\frac{1}{bt}} x^{2}
\end{equation}
where $\tilde{A}=\frac{A\omega_{D}^{4}}{v/L}$ and $\tilde{B}=\frac{B\Theta_{D}\omega_{D}^{2}}{v/L}$ are also dimensionless. Together, there are five fitting parameters for both $\kappa_{xx}(t)$ and $\kappa_{xy}(t)$: $\{\tilde{A},\tilde{B},b,\kappa_{0}, \bar{\tau}_{s}^{-1}\}$. 
For $\mathrm{La_2CuO_4}$ and $\mathrm{Sr_2CuO_2Cl_2}$, we take $v$=5.2~km/s~\cite{Suzuki2000} and $\Theta_D=385$~K~\cite{Ginsberg} for both LCO and SCOC (for lack of data on SCOC). The fitted parameters are:
\begin{align}
    P_{\text{La}} &= \{\tilde{A}=5.2\times10^{4},\tilde{B}=1.4\times10^{4},b=4.5,\kappa_{0}=6.8\times10^{4} \text{~W/(K m)} ,\bar{\tau}_{s}^{-1}=-7.6 \times 10^{-3}\} \\ 
    P_{\text{Sr}}&=\{\tilde{A}=9.5\times10^{3},\tilde{B}=6.5\times10^{3},b=5.0,\kappa_{0}=1.52\times10^{4} \text{~W/(K m)} ,\bar{\tau}_{s}^{-1}=-7.2\times 10^{-3} \}
\end{align}
This corresponds to the usual parameters
\begin{align}
 P_{La}&=\left\{ L=62 \mathrm{\mu m},A=6.7\times10^{-43} \mathrm{s^{3}},B=1.2\times10^{-18} \mathrm{s K^{-1}} ,b=4.5,\tau_{s}^{-1}=-0.64\times10^{6} \mathrm{s^{-1}} \right\}  \\
    P_{Sr}&=\left\{ L=14\mathrm{\mu m},A=5.6\times10^{-43} \mathrm{s^{3}},B=2.5\times10^{-18} \mathrm{s K^{-1}} ,b=5.0,\tau_{s}^{-1}=-2.7\times10^{6} \mathrm{s^{-1}} \right\} 
\end{align}
The resulting fits are shown in Fig.~\ref{fig:fullmodel}

\subsection{Approximation to the skew scattering model in intermediate temperatures}
The formidable integrals of $\kappa_{xx}$ and $\kappa_{xy}$ usually only admit a simple approximation in the low T or high T limit, which does not explain the linearity in our case.
In this section, we will perform an approximation in the intermediate temperature range 30-80~K. For a typical phonon of energy $\hbar \omega=k_B T$, we assume the magnitude of the three scattering rates to be
\begin{equation}
    \tau_{\text{b}}^{-1} \ll \tau_{\text{imp}}^{-1}(\omega) \sim \tau_{\text{Umk}}^{-1}(\omega,T)
\end{equation}
Under this assumption, we will show that $\kappa_{xx}$ and $\kappa_{xy}$ can be reduced to (\ref{eq:kinetic}). To start, the dimensionless scattering rates $\tilde{\tau}^{-1}(x,t)=1+\tilde{A}x^{4}t^{4}+\tilde{B}t^{3}x^{2}e^{-\frac{1}{bt}}$ can be put more conveniently by defining $\tilde{A}= a^{-4},\tilde{B}= c^{-2}$, so we have
\begin{equation}
    \tau^{-1}(x,t)=1+(tx/a)^{4}+t^{3}e^{-\frac{1}{bt}}(x/c)^{2}
\end{equation}
The temperature range 30-80~K corresponds to $t\in [1/13,1/5]$.
In assumption
$ \tau_{\text{b}}^{-1} \ll \tau_{\text{imp}}^{-1}$
we ignore the boundary scattering term for most phonons excited in this temperature range. 
However, we must also remove the low frequency modes dominated by boundary scattering, since they would otherwise have infinite conductivity.
The lower bound of the integral can be set by $(t x / a)^4 \sim 1$ and thus $x_{\min } \sim a / t$. Above this frequency, the scattering rate can be written as
\begin{equation}
    \tau^{-1}(x, t) \approx(t x / a)^4\left[1+\frac{a^4}{c^2} t^{-1} e^{-\frac{1}{b t}} x^{-2}\right]
\end{equation}
The two terms in brackets show the relative magnitude of impurity and Umklapp scatterings.
In the weak Umklapp limit $\tau_{\text{Umk}}^{-1} = 0$, $\kappa_{xx}$ would increase monotonically with $t$. 
On the other hand, in the strong Umklapp limit $\kappa_{x x} \sim e^{1 / b t}$ decay exponentially fast.
Therefore, we focus on an intermediate regime where $\tau_{\text{imp}}^{-1} \sim \tau_{\text{Umk}}^{-1}$, in which Umklapp scattering is comparable with impurity scattering but the $e^{1 / b t}$ factor has yet to set in.
Notice the tempearture dependence $t^{-1} e^{-\frac{1}{b t}}$ peaks at $t_u=1/b$ and is slow-varying in the range $[1/13, 1/5]$, so we can replace it by the peak value: $t_u^{-1} e^{-\frac{1}{b t_u}} = b / e$ and obtain a $t$-independent factor $[1+\frac{a^4 b}{c^2 e} x^{-2}]$.
At a given temperature $T$, a typical phonon has frequency $x = \hbar \omega / k_B T \sim 1$. We then require $\tau_{\text{imp}}^{-1} \sim \tau_{\text{Umk}}^{-1}$ to hold at this frequency, which leads to $\frac{a^4 b}{c^2 e} \sim 1$ and a scattering rate
\begin{equation}
    \tau^{-1}(x,t) \approx (tx/a)^{4}(1+x^{-2})
\end{equation}
Using this approximate scattering rate, and after simple manipulations, we have

\begin{align}
& \kappa_{x x} \approx \kappa_0 a^4 t^{-1} \int_{a / t}^{1 / t} d x F(x) \\
& \kappa_{x y} \approx \kappa_0 a^8 \bar{\tau}_s^{-1} t^{-5} \int_{a / t}^{1 / t} d x G(x)
\end{align}
\begin{equation}
    F(x)=J_0(x) \frac{x^2}{1+x^2}, \quad G(x)=J_0(x) \frac{1}{\left(1+x^2\right)^2}
\end{equation}
At large $x$, the integrand $F(x) \rightarrow e^{-x}$ and $G(x) \rightarrow x^{-4} e^{-x}$ decay rapidly, having a significant contribution only within $x<4$ and $x<2$. 
The upper bound $1 / t \geq 5$ can then be set to $\infty$.
We can approximation $F(x)$ within $x<4$ as
\begin{equation}
    F(x) \approx e^{-x}
\end{equation}
based on the observation that $F(x \rightarrow 0)=1$ and $F(x \rightarrow \infty)=e^{-x}$.
This gives us the final form for $\kappa_{x x}$ :
\begin{equation}
    \kappa_{x x} \approx \kappa_0 a^4 t^{-1} \int_{a / t}^{\infty} d x e^{-x} =\kappa_0 a^4 t^{-1} e^{-a / t}
\end{equation}
The approximation of $G(x)$ within $x<2$ is
\begin{equation}
    G(x)=F(x) \frac{1}{1+x^2} x^{-2} \approx \frac{e^{-x}}{1+x^2} x^{-2} \approx \frac{e}{2} e^{-2 x} x^{-2}
\end{equation}
where we used the same approximation for $F(x)$ in the first step. In the second step, we Taylor expand $\log(\frac{e^{-x}}{1+x^2})= -x-\log (1+x^2) \approx (-1-\log 2)-2(x-1)$ around the inflection point $x=1$. One can check that this function is indeed almost linear in $0.5<x<2$. 
We can now evaluate the $\kappa_{xy}$ integral
\begin{align}
\kappa_{x y} & \approx \frac{e}{2} \kappa_0 a^8 \bar{\tau}_s^{-1} t^{-5} \int_{a / t}^{\infty} d x e^{-2 x} x^{-2} \\
& =\frac{e}{2} a^8 \kappa_0 \bar{\tau}_s^{-1} t^{-5}\left(e^{-2a/t} \frac{t}{a}-2 \Gamma\left(0, 2a/t \right)\right) \\
& \approx \frac{e}{4} a^6 \kappa_0 \bar{\tau}_s^{-1} t^{-3} e^{-2a/t}
\end{align}
where in the last step we Taylor expanded the incomplete gamma function $\Gamma(s, x)$ to lowest order around $t=0$ (since $t<1/5$). 
Identifying the parameters $\tilde{t}=t/a=\frac{T}{a\Theta_{D}}$ (which is $t$ in (\ref{eq:kinetic})), $\kappa_{L}=\kappa_{0}a^{3}$ and $\tilde{\tau}_{s}=\frac{e}{4}\bar{\tau}_{s}^{-1}=-\frac{e}{4}\frac{\tau_{s}^{-1}}{v/L}$, we arrive at (\ref{eq:kinetic}). 
We found a set of parameters for LCO and SCOC: 
\begin{align}
P_{La} & =\{\kappa_{L}=37 \text{~W/(K m)},\quad \tilde{\tau}_{s}^{-1}=-6.3\times10^{-3}, \quad a=0.085\}\\
P_{Sr} & =\{\kappa_{L}=23 \text{~W/(K m)},\quad \tilde{\tau}_{s}^{-1}=-5.9\times10^{-3}, \quad a=0.091\}
\end{align}
which can also be converted to
\begin{align}
P_{La} & =\{L=55~\mathrm{\mu m},\quad A=2.9\times10^{-43}~\mathrm{s^3},\tau_{s}^{-1}=-0.88\times10^{6}\mathrm{s^{-1}}\}\\
P_{Sr} & =\{L=27~\mathrm{\mu m},\quad A=4.2\times10^{-43}~\mathrm{s^3},\tau_{s}^{-1}=-1.7\times10^{6}\mathrm{s^{-1}}\}
\end{align}
where we use $a/0.8$ instead of $a$ in the conversion, since ignoring boundary scattering will lead to an overestimation of impurity scattering rate.

\section{Thermal conductivity data for additional cuprates}
\begin{figure}[h]
\centering
\includegraphics[scale=0.5]{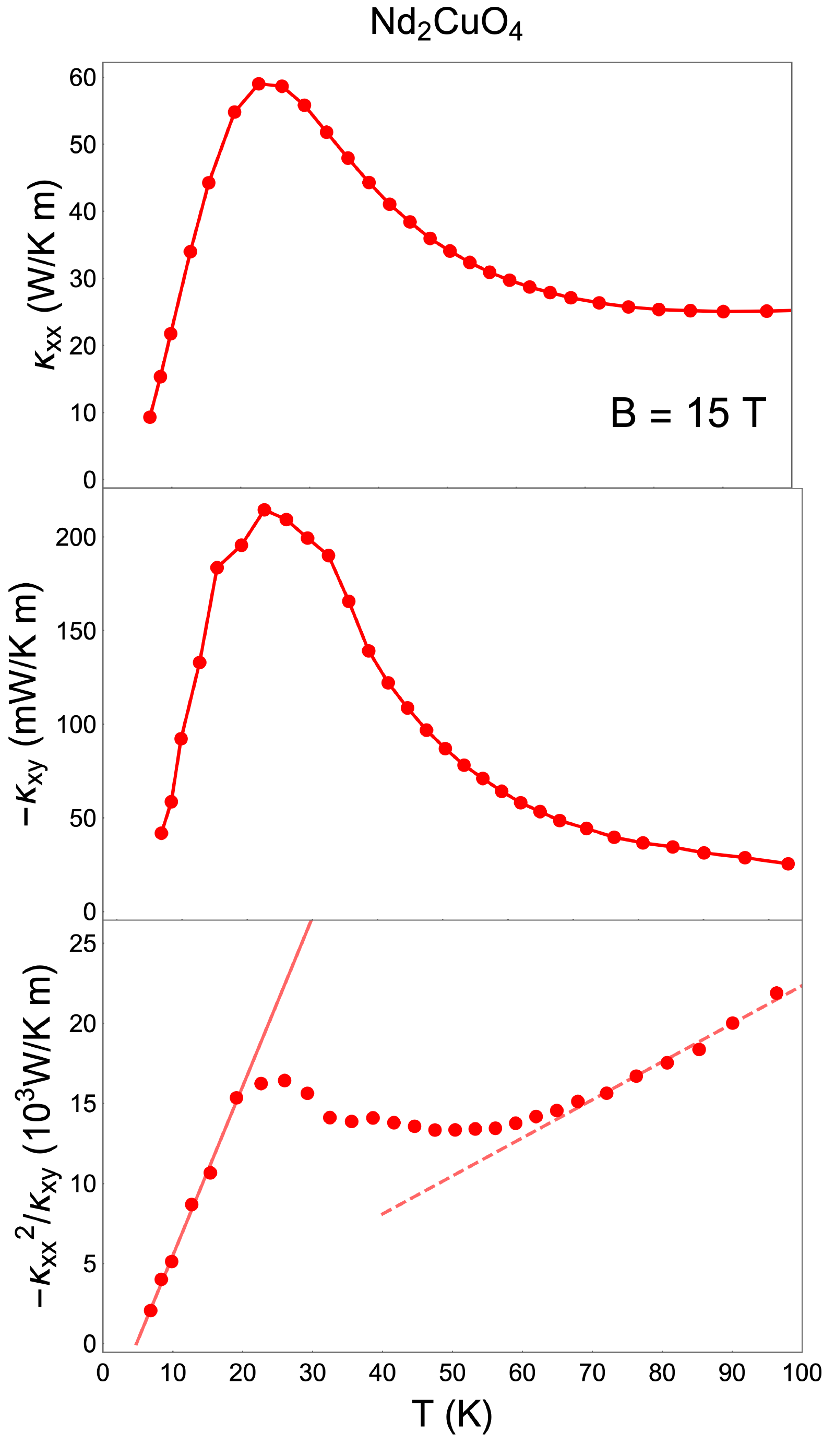}
\caption{\label{fig:Nd2CuO4} 
Thermal conductivities (top and middle), and $\rho_H^{-1}$ versus temperature for undoped cuprate Mott insulator $\mathrm{Nd_2CuO_4}$. 
We see a deviation from linearity between $20$ and $\mbox{70 K}$.
The low-$T$ linear fit (solid line) corresponds to $A=1.1 \mathrm{\times 10^3 W/(K^2 m)}$ and $A_0=-5.0 \mathrm{\times 10^3 W/(K m)}$, 
while the high-$T$ linear fit (dashed line) corresponds to $A=0.24 \mathrm{\times 10^3 W/(K^2 m)}$ and $A_0=-1.4 \mathrm{\times 10^3 W/(K m)}$.
Data from~\cite{Boulanger2020}.
}
\end{figure}

{\bf Mott insulating NCO}---
$\mathrm{Nd_2CuO_4}$ is another undoped cuprate Mott insulator with a tetragonal crystal structure.  
Similar to LCO and SCOC, $\mathrm{Nd_2CuO_4}$ has quasi-2D antiferromagnetic order, with Néel temperature $T_N=245$~K~\cite{Chattopadhyay1991}. 
The bottom panel of Fig.~\ref{fig:Nd2CuO4} shows a clear deviation from linearity between $20-70$~K, which roughly corresponds to the spin reorientation range $30-85$~K. 
Further analysis is required to explain this deviation.

\newpage
\textbf{Electron-doped cuprate PCCO}---
$\mathrm{Pr}_{2-x}\mathrm{Ce}_x\mathrm{CuO}_4$ (PCCO) at doping $x=0.15$ satisfies the linear relation $\rho_H^{-1}=AT+A_0$ 
in the range $5-40$~K with the same slope $A$ as NCCO at doping $x=0.17$ (Fig.~\ref{fig:NCCO}). 
The non-linear dependence starting at $40$~K roughly corresponds with the onset of an additional contribution to $\kappa_{xx}$.

\begin{figure}[h]
\centering
\includegraphics[scale=0.5]{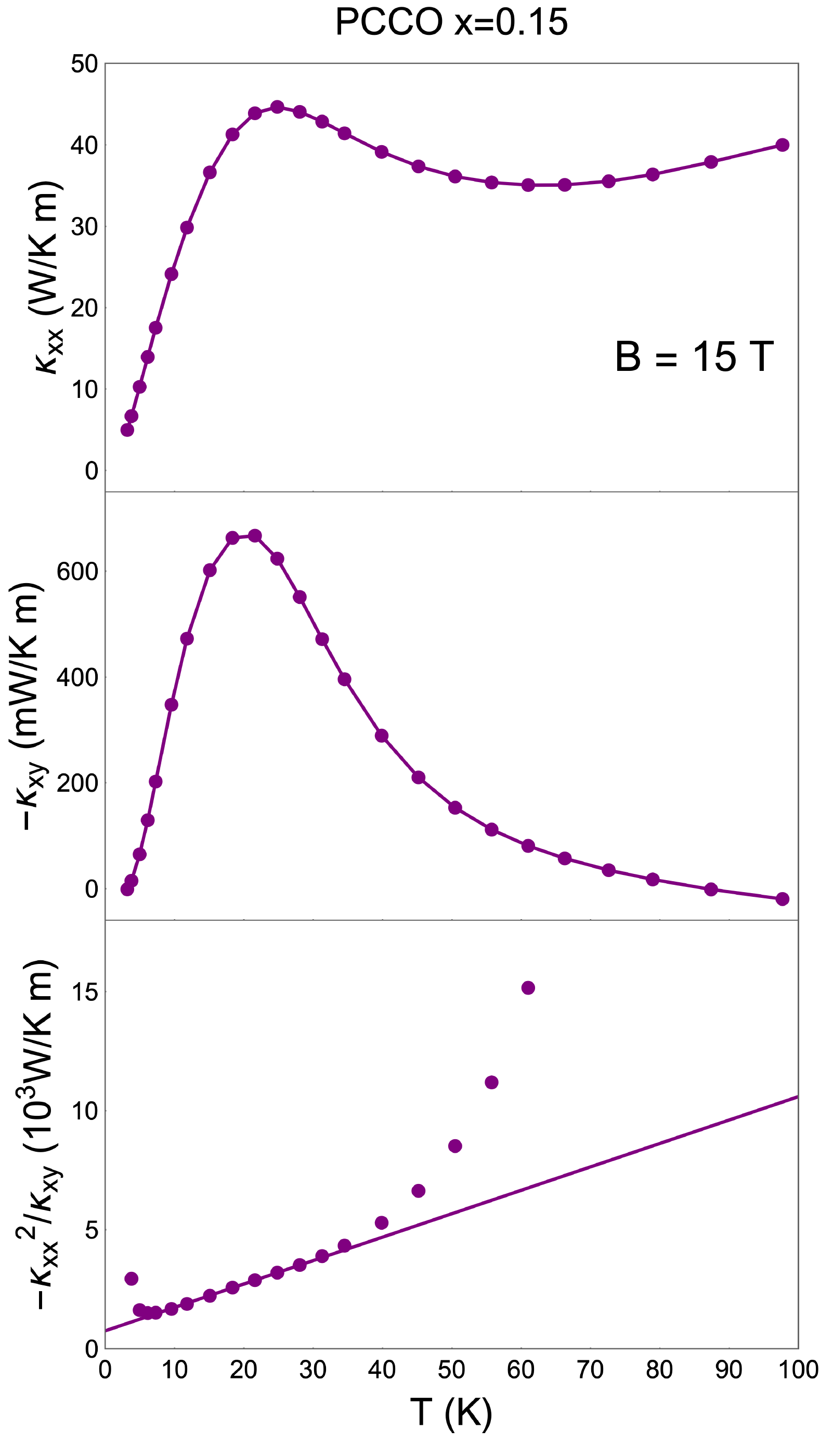}
\caption{\label{fig:PCCO} 
Thermal conductivities (top and middle), and $\rho_H^{-1}$ versus temperature for electron-doped cuprate $\mathrm{Pr_{2-x}Ce_xCuO_4}$ (PCCO). 
We see a deviation from linearity at around $40$~K. 
The linear fit corresponds to $A=0.10 \mathrm{\times 10^3 W/(K^2 m)}$ and $A_0=0.67 \mathrm{\times 10^3 W/(K m)}$.
Data from~\cite{Boulanger2022}.
}
\end{figure}

\end{document}